% hep-th version v2
\input harvmac
%\input epsf
%\draftmode

\overfullrule=0pt
\abovedisplayskip=12pt plus 4pt minus 0pt
\belowdisplayskip=12pt plus 4pt minus 0pt
%macros
%
\def\tilde{\widetilde}
\def\bar{\overline}

\def\tC{{\tilde C}}

\def\tF{{\tilde F}}
\def\tSigma{{\tilde \Sigma}}

\def\bx{{\bf x}}
\def\bX{{\bf X}}
\def\bigone{\hbox{1\kern -.23em {\rm l}}}
\def\ZZ{\hbox{\zfont Z\kern-.4emZ}}
\def\half{{\litfont {1 \over 2}}}
\def\cO{{\cal O}}

\def\tr{{\rm tr}\,}

\def\Str{{\rm Str}\,}
\def\Pf{{\rm Pf}\,}
\def\hA{{\hat A}}
\def\hF{{\hat F}}
\def\hT{{\hat T}}
\def\hPhi{{\hat \Phi}}

\font\litfont=cmr6

\def\cQ{{\cal Q}}

\def\ola{\overleftarrow}
\def\ora{\overrightarrow}

\lref\myers{R. C. Myers, {\it ``Dielectric-Branes''}, hep-th/9910053,
JHEP {\bf 12} (1999) 022.}
\lref\ramswati{M. Van Raamsdonk and W. Taylor, {\it ``Multiple
Dp-branes in Weak Background Fields''}, hep-th/9910052,
Nucl. Phys. {\bf B573} (2000) 703.}
\lref\watirev{W. Taylor, {\it ``The M(atrix) Model of M Theory''},
hep-th/0002016.}
\lref\seiwit{N. Seiberg and E. Witten, {\it ``String Theory and
Noncommutative Geometry''}, hep-th/9908142, JHEP {\bf 09} (1999) 032.}
\lref\seibnew{N. Seiberg, {\it ``A Note on Background
Independence in Noncommutative Gauge Theories, Matrix Model and
Tachyon Condensation''}, hep-th/0008013.}
\lref\twosens{B. Janssen and P. Meessen, {\it ``A Nonabelian
Chern-Simons Term for Non-BPS D-Branes''}, hep-th/0009025.}
\lref\garousitach{M.R. Garousi, {\it ``Tachyon Couplings on Non-BPS
D-Branes and Dirac-Born-Infeld Action''}, hep-th/0003122,
Nucl. Phys. {\bf B584} (2000) 284.}
\lref\smnvs{S. Mukhi and N.V. Suryanarayana, {\it ``Chern-Simons 
Terms on Noncommutative Branes''}, hep-th/0009101, 
JHEP {\bf 11} (2000) 006.}
\lref\tatar{R. Tatar, {\it ``T-Duality and Actions for Noncommutative
D-branes''}, hep-th/0011057.}
\lref\iikk{N. Ishibashi, S. Iso, H. Kawai and Y. Kitazawa, {\it
``Wilson Loops in Noncommutative Yang-Mills''},
hep-th/9910004, Nucl. Phys. {\bf B573} (2000) 573.}
\lref\reyunge{S.-J. Rey and R. von Unge, {\it ``S-duality,
Noncritical Open String and Noncommutative Gauge Theory''},
hep-th/0007089, Phys. Lett. {\bf B499} (2001) 215.}
\lref\dasrey{S.R. Das and S.-J. Rey, {\it ``Open Wilson Lines
in Noncommutative Gauge Theory and Tomography of Holographic Dual
Supergravity''}, hep-th/0008042, Nucl. Phys. {\bf B590} (2000) 453.}
\lref\ghi{D.J. Gross, A. Hashimoto and N. Itzhaki, {\it ``Observables 
of Noncommutative Gauge Theories''}, hep-th/0008075.}
\lref\garousi{M. Garousi, {\it ``Noncommutative World Volume
Interactions on D-branes and Dirac-Born-Infeld Action''},
hep-th/9909214, Nucl. Phys. {\bf B579} (2000) 209.}
\lref\mehenwise{T. Mehen and M. Wise, {\it ``Generalized $*$-products, 
Wilson Lines and the Solution of the Seiberg-Witten Equations''},
hep-th/0010204, JHEP {\bf 12} (2000) 008.}
\lref\micheliu{H. Liu and J. Michelson, {\it ``~$*$-Trek: The One
Loop N=4 Noncommutative SYM Action''}, hep-th/0008205.}
\lref\liu{H. Liu, {\it ``~$*$-Trek II: $*_n$ Operations, Open
Wilson Lines and the Seiberg-Witten Map''}, hep-th/0011125.}
\lref\dastrivedi{S.R. Das and S. Trivedi, {\it ``Supergravity
Couplings to Noncommutative Branes, Open Wilson Lines and Generalized
Star Products''}, hep-th/0011131, JHEP {\bf 02} (2001) 046.}
\lref\wyllard{N. Wyllard, {\it ``Derivative Corrections to D-brane 
Actions with Constant Background  Fields''}, hep-th/0008125,
Nucl. Phys. {\bf B598} (2001) 247.}
\lref\kontsevich{M. Kontsevich, {\it ``Deformation Quantization of
Poisson Manifolds I''}, q-alg/9709040.}
\lref\catfeld{A.S. Cattaneo and G. Felder, {\it ``A Path Integral
Approach to the Kontsevich Quantization Formula''}, math.QA/9902090,
Comm. Math. Phys. {\bf 212} (2000) 591.}
\lref\jurcoschupp{B. Jurco and P. Schupp, {\it ``Noncommutative 
Yang-Mills from Equivalence of Star Products''},
hep-th/0001032, Eur. Phys. J. {\bf C14} (2000) 367.}
\lref\krs{P. Kraus, A. Rajaraman and S. Shenker,
{\it ``Tachyon Condensation in Noncommutative Gauge Theory''},
hep-th/0010016, Nucl. Phys. {\bf B598} (2001) 169.}
\lref\mandalwadia{G. Mandal and S.R. Wadia, 
{\it ``Matrix Model, Noncommutative Gauge Theory and the 
Tachyon Potential''}, hep-th/0011094, Nucl. Phys. {\bf B599}
(2001) 137.}
\lref\kenwilk{C. Kennedy and A. Wilkins, {\it ``Ramond-Ramond 
Couplings on Brane-Antibrane Systems''}, hep-th/9905195, 
Phys. Lett. {\bf B464} (1999) 206.}
\lref\kral{P. Kraus and F. Larsen, {\it ``Boundary String Field 
Theory of the $D{\bar D}$ System''}, hep-th/0012198.}
\lref\ttu{T. Takayanagi, S. Terashima and T. Uesugi, {\it 
``Brane-Antibrane Action from Boundary String Field Theory''},
hep-th/0012210.}
\lref\miao{M. Li, {\it ``Note on Noncommutative Tachyon in Matrix 
Models''}, hep-th/0010058.}
\lref\okawao{Y. Okawa and H. Ooguri, {\it ``An Exact Solution to 
Seiberg-Witten Equation of Noncommutative Gauge Theory''},
hep-th/0104036.}
\lref\kmm{D. Kutasov, M. Marino and G. Moore, 
{\it ``Remarks on Tachyon Condensation in Superstring Field Theory''},
hep-th/0010108.}
\lref\berle{D. Berenstein and R. Leigh, {\it ``Observations on 
Non-commutative Field Theories in Coordinate Space''}, 
hep-th/0102158.}

{\nopagenumbers
\Title{\vbox{
\hbox{hep-th/0104045}
\hbox{TIFR/TH/01-13}
\hbox{NSF-ITP-01-26}}}
{\vtop{
\centerline{Gauge-Invariant Couplings of  Noncommutative}
\medskip
\centerline{Branes to Ramond-Ramond Backgrounds}}}
\centerline{{Sunil Mukhi$^{1,2}$}\footnote{}{E-mail: mukhi@tifr.res.in,
nemani@tifr.res.in} and {Nemani V. Suryanarayana$^1$}}
\vskip 8pt
\centerline{\it $\!\!^1$ Tata Institute of Fundamental Research}
\centerline{\it Homi Bhabha Rd, Mumbai 400 005, India}
\vskip8pt

\centerline{\it $\!\!^2$ Institute for Theoretical Physics}
\centerline{\it University of California, Santa Barbara, CA 93106, 
U.S.A.}

\vskip 1.5truecm
\centerline{\bf ABSTRACT}
\medskip
We derive the couplings of noncommutative D-branes to spatially
varying Ramond-Ramond fields, extending our earlier results in
hep-th/0009101. These couplings are expressed in terms of $*_n$
products of operators involving open Wilson lines. Equivalence of the
noncommutative to the commutative couplings implies interesting
identities as well as an expression for the Seiberg-Witten map that
was previously conjectured. We generalise our couplings to include
transverse scalars, thereby obtaining a Seiberg-Witten map relating
commutative and noncommutative descriptions of these scalars.  RR
couplings for unstable non-BPS branes are also proposed.

\vfill
\Date{April 2001}
\eject}
\ftno=0

\listtoc
\writetoc

\newsec{Introduction}

In a previous paper\refs\smnvs, we examined the Chern-Simons terms on
noncommutative branes in the ``background independent'' description
$\Phi = - B$\refs{\seiwit,\seibnew}. These terms describe the
couplings of D-branes to Ramond-Ramond (RR) gauge potentials of
various form degrees. In that work, the noncommutative description for
D-branes with a B-field was given for both BPS and non-BPS branes,
with the restriction that the RR potentials to which they couple are
constant in space. It was shown subsequently\refs\tatar\ that our
expressions can be rederived using the fact that D-branes on a torus
with a B-field turn into slanted branes under T-duality.

The noncommutative couplings of Ref.\refs\smnvs\ were shown to be
consistent with the interpretation that a noncommutative brane is a
configuration of infinitely many lower dimensional branes. In
particular, the well-known terms which couple RR backgrounds to
non-abelian scalars on a collection of BPS
D-branes\refs{\myers,\ramswati,\watirev} were
reproduced. Generalisations of these terms to unstable non-BPS
D-branes were also obtained and compared with results in
Ref.\refs\twosens.

In the present paper, our goal is to extend these couplings to
nonconstant RR backgrounds. We will propose formulae for these
couplings, which involve the open Wilson lines (``Wilson tails'') that
have recently attracted much interest in the study of noncommutative
field theory\refs{\iikk,\reyunge,\dasrey,\ghi}. The generalized $*'$
and $*_n$ products that have been discussed in Refs.\refs{\garousi,
\mehenwise,\micheliu,\liu,\dastrivedi} will play an important role
\foot{A recently developed approach to write smeared operators and 
compute correlation functions directly in coordinate space\refs\berle\
might provide a useful alternative framework for this kind of
investigation.}.

This investigation is of interest first of all because RR couplings
characterise much of the physical behaviour of D-branes in superstring
theory, both BPS and non-BPS. Hence it is important to know how these
couplings generalise to the noncommutative setting. Additionally, we
will find that the equivalence of noncommutative couplings to
commutative ones gives rise to a variety of interesting
identities. Some of these, which are novel, amount to nontrivial
properties of the $*$ product. We will prove these identities
explicitly in special cases, although one can also turn the logic
around and claim that the new identities follow from the equivalence
of commutative and noncommutative RR couplings of D-branes. Other
identities that we will encounter are actually maps from
noncommutative fields to their commutative counterparts. These embody
the change of variables between these two types of fields, known as
the Seiberg-Witten map. Again, one can try to confirm the expressions
for the Seiberg-Witten map that we find (one of them was conjectured
earlier by Liu\refs\liu) from independent computations. Alternatively
one can claim that they follow from equivalence of commutative and
noncommutative RR couplings. It therefore emerges that this
equivalence is a powerful tool to derive properties of noncommutative
field theory.  Finally, a third type of identity that we will come
across holds only in the DBI approximation of slowly-varying
fields. In this case it will be interesting to examine how the
derivative corrections are incorporated, a point to which we hope to
return in the future.

When this manuscript was nearly complete, we learned of forthcoming
papers by H. Liu and J. Michelson, and by Y. Okawa and
H. Ooguri\refs\okawao, having substantial overlap with our work. We are
grateful to Jeremy Michelson and Hirosi Ooguri for informing us about
their results.

\newsec{Non-constant RR fields}

In this section we write down the couplings of a noncommutative brane
to spatially varying Ramond-Ramond potentials. We will neglect
transverse scalars, to which we return in the following section. Thus
the results of this section hold only when the transverse scalars are
set to zero, alternatively they hold in complete generality for a
Euclidean D9-brane.

One may be concerned that the RR couplings of a commutative D-brane
can have derivative corrections when the RR field is spatially
varying. However, the coupling of a D$p$-brane to the top form
$C^{p+1}$ and to the next lower form $C^{p-1}$ are exactly known in
the commutative case, because for these two it has been
argued\refs\wyllard\ that derivative corrections are absent. For the
other RR forms, the derivative corrections are only partially
determined, and in these cases our results for noncommutative
couplings should agree with the standard commutative couplings only
for slowly-varying fields.

\subsec{Coupling to the RR Top-Form $C^{p+1}$}

Let us consider a noncommutative Euclidean D$p$-brane with an even
number $p+1$ of world-volume directions. As is well-known, in a
noncommutative theory one does not have gauge-invariant local
operators, because of the non-locality induced by the
noncommutativity. One does, however, have gauge-invariant operators of
definite momentum. Hence, in order to study the coupling of spatially
varying RR gauge potentials to noncommutative branes, we will choose
the RR potential to be evaluated at a definite momentum $k^i$. The
RR couplings to noncommutative branes for $k^i =0$ were found in
Ref.\refs\smnvs.

Now we will use the following important results from
Refs.\refs{\liu,\dastrivedi}. Given a collection of local operators
$\cO_I(x)$ on the brane world-volume which transform in the adjoint
under gauge transformations, one can obtain a natural gauge-invariant
operator of fixed momentum $k^i$ by smearing the locations of these
operators along a straight contour given by $\xi^i(\tau) =
\theta^{ij}k_j\,\tau$ with $0\le\tau\le 1$, and multiplying the
product by a Wilson line $W(x,C)$ along the same contour,
\eqn\wilsonline{
W(x,C) \equiv \exp\left( i \int_0^1 d\tau {\del\xi^i(\tau)\over
\del\tau}\hA_i\Big(x+ \xi(\tau)\Big) \right) }
Here and in what follows, $\hA$ denotes the noncommutative gauge
field, and $\hF$ the corresponding field strength. 

The resulting formula for the gauge-invariant operator is:
\eqn\ginvprod{
\eqalign{
\cQ(k) &= \int { d^{p+1}x\over (2\pi)^{p+1}} \left( \prod_{I=1}^n 
\int_0^1 d\tau_I \right)
P_*\left[ W(x,C) \prod_{I=1}^n \cO_I\Big(x+\xi(\tau_I)\Big)\right] *
e^{ik.x} \cr &= \int { d^{p+1}x\over (2\pi)^{p+1}}~ L_* \left[ W(x,C)
\prod_{I=1}^n \cO_I(x)\right] * e^{ik.x}\cr }}
where $P_*$ denotes path-ordering with respect to the $*$-product,
while $L_*$ is an abbreviation for the combined path-ordering and
integrations over $\tau_I$. In this formula the operators ${\cal O}_I$
are smeared over the straight contour of the Wilson line. This
prescription arises by starting with the symmetrised-trace action for
infinitely many D-instantons and expanding it around  the
configuration  describing a noncommutative D$p$-brane.

Expanding the Wilson line and using manipulations described in  
Refs.\refs{\mehenwise,\liu,\dastrivedi}, we get
\eqn\liutwo{
\cQ(k)= \sum_{m=0}^{\infty}\int { d^{p+1}x\over (2\pi)^{p+1}} ~ 
\cQ_m(x) \, e^{ik.x} }
where
\eqn\liumore{
\cQ_m(x) = {1\over m!}(\theta\partial)^{i_1}....(\theta\partial)^{i_m}
\langle {\cal O}_1(x),..., {\cal O}_n(x), {\hat A}_{i_1}(x),..., 
{\hat A}_{i_m}(x)\rangle_{*_{m+n}}}
Here we have introduced the notation $\langle
f_1(x),f_2(x),...,f_p(x)\rangle_{*_p}$ for the $*_p$ product of $p$
functions, as defined for example in the appendix of Ref.\liu.  We
note here the simple formula for $*_2$:
\eqn\startwodef{
\langle f(x),g(x) \rangle_{*_2} \equiv 
f(x){\sin(\half\ola{\partial_p}\theta^{pq}\ora{\partial_q})
\over \half\ola{\partial_p}\theta^{pq}\ora{\partial_q} } g(x) }

The above procedure for defining gauge-invariant operators is useful
when applied to couplings between closed-string and open-string modes,
namely couplings of a noncommutative D-brane to bulk fields. Once we
know the coupling of a generic closed-string supergravity Fourier
mode\foot{Fourier modes will always be denoted with a tilde.}
$\tSigma(k)$ to a D-brane for $k=0$, then we can derive the coupling
at nonzero momentum by suitably inserting an open Wilson line as
above. More precisely, if the zero-momentum coupling is
\eqn\zerocoup{
\tSigma(0)\, \int { d^{p+1}x\over (2\pi)^{p+1}}\, {1\over\Pf\theta} 
\cO_\Sigma(A,X) }
where $A$ is the gauge field and $X$ are the transverse scalars, then
the coupling at nonzero momentum is given by
\eqn\nonccoup{
\tSigma(-k) \int { d^{p+1}x\over (2\pi)^{p+1}}\, {1\over \Pf\theta}\, 
L_* \Big[ \cO_\Sigma(A,X)\, W(x,C)\Big]* e^{ik.x} }
The constant factor $\Pf\theta\equiv\sqrt{\det\theta^{ij}}$ has been
written explicitly, instead of absorbing it into the definition of
$\cO_\Sigma$, for later convenience.

In our case the relevant closed string mode is the RR gauge potential
$\tC^{p+1}(k)$, and, as shown in Ref.\refs\smnvs, the role of
$\cO_\Sigma$ is played by the operator $\mu_p\, \Pf Q$ where $\mu_p$ is
the brane tension and
\eqn\qinv{
Q^{ij} \equiv \theta^{ij}-\theta^{ik}
{\hat F}_{kl}\,\theta^{lj}}

Hence we deduce the coupling of this brane to the form
$C^{(p+1)}$, in momentum space, to be:
\eqn\csone{
\mu_p\,\epsilon^{i_1...i_{p+1}}\,
{\tilde C}^{(p+1)}_{i_1...i_{p+1}}(-k)\int
{ d^{p+1}x\over (2\pi)^{p+1}} 
~L_* \Big[{\sqrt{\hbox{det}(1-\theta \hat F)}}~ W(x,
C)\Big] * e^{ik.x}}
On the other hand, we know that the coupling of a D$p$-brane to a
$C^{(p+1)}$ form is given in the commutative description by
\eqn\comcsone{
\mu_p \,\epsilon^{i_1...i_{p+1}} \, {\tilde
C}^{(p+1)}_{i_1...i_{p+1}}(-k) ~ \delta^{(p+1)}(k)}

As equations \csone\ and \comcsone\ describe the same system in two
different descriptions, they must be equal. Thus we predict the
identity\foot{The idea that this identity should hold arose in
discussions with Sumit Das.}:
\eqn\idone{
\delta^{(p+1)}(k) = \int { d^{p+1}x\over (2\pi)^{p+1}}~ L_{*}
\big[{\sqrt{\hbox{det}(1-\theta \hat F)}}~ 
W(x, C)\big] * e^{ik.x}}
This amounts to saying that the right hand side is actually
independent of $\hA$, a rather nontrivial fact.

Let us show explicitly that this identity holds to order ${\cal
O}(\hat A^3)$ and to all orders in $\theta$. First, the operator
multiplying the open Wilson line is expanded as:
\eqn\opexpan{
\eqalign{
\sqrt{\hbox{det}(1 - \theta\hat F)} =& 1 - {1\over2}\tr (\theta\hat F) 
- {1\over4}\tr (\theta\hat F \theta\hat F) 
+ {1\over8}(\tr (\theta\hat F))^2  \cr
&- {1\over 48}(\tr (\theta\hat F))^3 + {1\over 8}
(\tr \theta\hat F)\tr (\theta\hat F\theta\hat F) 
- {1\over 6}\tr (\theta\hat F\theta\hat F\theta\hat F)+ {\cal
O}(\hat F^4)}}
Then the first four terms in $\cQ(x)$ (Eq.\liutwo) can be explicitly
computed, giving:
\eqn\qs{
\eqalign{
\cQ_0(x) =&~ 1 - {1\over 2} \theta^{ij}\hat F_{ji} - {1\over 4}
\theta^{ij}\theta^{kl}\langle \hat F_{jk}, \hat
F_{li}\rangle_{*_2} + {1\over 8}\theta^{ij}\theta^{kl} \langle
\hat F_{ji}, \hat F_{lk}\rangle_{*_2} \cr
& - {1\over 6}\theta^{ij}\theta^{kl}\theta^{mn}\langle \hat F_{jk},
\hat F_{lm}, \hat F_{ni}\rangle_{*_3} 
+ {1\over8}\theta^{ij}\theta^{kl}\theta^{mn}\langle \hat F_{ji},
\hat F_{lm}, \hat F_{nk}\rangle_{*_3} \cr
& - {1\over 48}\theta^{ij}\theta^{kl}\theta^{mn}\langle \hat F_{ji},
\hat F_{lk}, \hat F_{nm}\rangle_{*_3} + {\cal O}(\hat A^4)\cr
\cQ_1(x) =&~ \theta^{ij}\partial_j\hat A_i - {1\over
2}\theta^{ij}\theta^{kl}\partial_l\langle \hat F_{ji}, \hat
A_k\rangle_{*_2} \cr
&- {1\over 4}\theta^{ij}\theta^{kl}\theta^{mn}
\partial_n\langle \hat F_{jk}, \hat F_{li}, \hat A_m\rangle_{*_3}  
+ {1\over 8}\theta^{ij}\theta^{kl}\theta^{mn} \partial_n\langle 
\hat F_{ji},
\hat F_{lk}, \hat A_m\rangle_{*_3}+ {\cal O}(\hat A^4)\cr
\cQ_2(x) =&~ {1\over2}\theta^{ij}\theta^{kl}\partial_j \partial_l\langle
\hat A_i, \hat A_k\rangle_{*_2} - {1\over 4}
\theta^{ij}\theta^{kl}\theta^{mn}\partial_l\partial_n\langle \hat
F_{ji}, \hat A_{k}, \hat A_m\rangle_{*_3}+ {\cal O}(\hat A^4)\cr
\cQ_3(x) =&~ {1\over
6}\theta^{ij}\theta^{kl}\theta^{mn}\partial_j\partial_l\partial_n\langle\hat
A_i, \hat A_k, \hat A_m\rangle_{*_3}+ {\cal O}(\hat A^4)}}
It is a straightforward, though lengthy, exercise to check that
\eqn\sumqs{
\cQ_0(x) + \cQ_1(x) + \cQ_2(x) + \cQ_3(x) = 1 + {\cal O}(\hat A^4)}
This proves our identity Eq.\idone, expressing the equivalence of the
coupling in Eq.\csone\ to that in Eq.\comcsone, up to ${\cal
O}(\hat A^4)$ terms.

In fact, we can prove that this identity holds to all orders in $\hat A$
for the special case where the noncommutativity parameter $\theta$ is
of rank two. In this case one can write
\eqn\ranktwo{
\sqrt{\det(1-\theta \hF)} = 1 + \theta^{12}\hF_{12} }
which is a considerable simplification of one factor in the formula.
The proof of Eq.\idone\ for the rank two case is given in 
Appendix A. At present we do not have an explicit proof in the most
general case.

It is illuminating to express Eq.\idone\ in the operator
formalism. We use the fact that
\eqn\inttr{
\int {d^{p+1}x\over (2\pi)^{p+1\over2}}\, {1\over \Pf\theta} 
\rightarrow \tr}
to rewrite the LHS of Eq.\idone\ as follows:
\eqn\lhsidone{
\eqalign{
\delta^{p+1}(k) &= \int\, {d^{p+1}x\over (2\pi)^{p+1}}\, e^{ik.x}\cr
&= {1\over (2\pi)^{p+1\over 2}}\, \tr (\Pf\theta~ e^{ik.\bx}) }}
The RHS of Eq.\idone\ can be converted to a symmetrised trace
involving $\bX^i\equiv \bx^i + \theta^{ij}\hA_j(\bx)$, following arguments 
in Ref.\dastrivedi, and it becomes:
\eqn\rhsidone{
{1\over (2\pi)^{p+1\over 2}}\, \Str (\Pf Q~ e^{ik.\bX}) }
where $\Str$ denotes the symmetrised trace. Finally, we use $[\bx^i,
\bx^j]=i\theta^{ij}$ and $[\bX^i, \bX^j]=iQ^{ij}(\bx)$. Then,
Eq.\idone\ takes the elegant form:
\eqn\idoneop{
\tr\left( \Pf[\bx^i,\bx^j]~ e^{ik.\bx}\right) = 
\Str\left( \Pf[\bX^i,\bX^j]~ e^{ik.\bX}\right) }
In this form, it is easy to see that Eq.\idone\ holds for constant
$\hF$, or equivalently for constant $Q$. In this special case it can
be proved by pulling $\Pf Q$ out of the symmetrised trace on the RHS,
and then using Eq.\inttr\ with $\theta$ replaced by $Q$.

For the more general case where $\hF$ and therefore $Q$ is spatially
varying, a suggestive line of argument runs as follows. To prove this
identity, we basically need to make the replacement:
\eqn\intstr{
\Str \Pf Q(X) \rightarrow \int {d^{p+1}X\over (2\pi)^{p+1\over2}} }
with the integrand on the RHS involving some suitable generalization
of the Moyal $*$ product. For constant $Q^{ij}$ this is valid with the
usual Moyal $*$ product, while for varying $Q^{ij}$ it requires the
deformation quantization of a Poisson structure with variable
coefficients. The existence of such a quantization is in fact
guaranteed by the work of Kontsevich\refs\kontsevich\ (see also
Ref.\refs{\catfeld,\jurcoschupp}), and was noted more recently in a
context similar to the present one in Ref.\liu. Now given such a
quantisation and its associated $*$ product, the RHS of Eq.\idoneop\
reduces to the integral over $X$ of a simple exponential $e^{ik.X}$,
and the result will presumably be a delta-function with any reasonable
$*$ product. It is important to find a rigorous proof of Eq.\idoneop\
along these lines, and also to investigate whether the generalised
associative product associated to $Q^{ij}(x)$ is helpful in writing
down noncommutative brane couplings.

\subsec{Coupling to the RR Form $C^{p-1}$ and the Seiberg-Witten Map}

Next let us turn to the coupling of a noncommutative $p$-brane to the
RR form $C^{(p-1)}$. In the commutative case this form appears in a
wedge product with the 2-form $B+F$. For the noncommutative brane in a
constant RR background, it was observed in Ref.\smnvs\ that $B+F$ must
be replaced by the 2-form $Q^{-1}$ with $Q^{ij}$ given by Eq.\qinv. It
follows that the coupling in the general noncommutative case (with
varying $C^{(p-1)}$) is:
\eqn\cstwo{
\mu_p\,\epsilon^{i_1...i_{p+1}}\,
{\tilde C}^{(p-1)}_{i_1...i_{p-1}}(-k)\int
{d^{p+1}x\over (2\pi)^{p+1}} \,
L_*\Big[{\sqrt{\hbox{det}(1-\theta \hat F)}}
\,(Q^{-1})_{i_pi_{p+1}} W(x, C)\Big] * e^{ik.x}}
For comparison, the coupling of a D$p$-brane to the form $C^{(p-1)}$ 
in terms of commutative variables is given by:
\eqn\comcstwo{
\eqalign{
\mu_p \, \epsilon^{i_1...i_{p+1}} \, {\tilde
C}^{(p-1)}_{i_1...i_{p-1}}(-k)  \int {d^{p+1}x\over (2\pi)^{p+1}}
~(B + F)_{i_pi_{p+1}}(x) 
\, e^{ik.x}\cr
=~ \mu_p \, \epsilon^{i_1...i_{p+1}} \, {\tilde
C}^{(p-1)}_{i_1...i_{p-1}}(-k)\,\Big[\delta^{p+1}(k)B_{i_pi_{p+1}} 
+ \tilde F_{i_pi_{p+1}}(k)\Big]}}
Next, rewrite $Q^{-1}$ as
\eqn\qinv{
\eqalign{
Q^{-1} &=~ \theta^{-1}\left[ 1+\theta\hat F(1-\theta\hat
F)^{-1}\right] \cr
&=~ B + \hat F(1-\theta\hat F)^{-1}}}
where we have used the relation $B = \theta^{-1}$. Using this relation 
and also Eq.\idone, we can rewrite Eq.\cstwo\ as:
\eqn\cstwoagn{
\eqalign{
&\mu_p\, \epsilon^{i_1...i_{p+1}}\,{\tilde
C}^{(p-1)}_{i_1...i_{p-1}}(-k)~
\Bigg\{\delta^{p+1}(k)\,B_{i_pi_{p+1}} +\cr 
& \int {d^{p+1}x\over (2\pi)^{p+1}} 
~L_*\Big[{\sqrt{\hbox{det}(1-\theta \hat F)}}\,
\big(\hF(1-\theta\hF)^{-1}\big)_{i_pi_{p+1}}\, W(x, C)\Big] *
e^{ik.x} \Bigg\}
}}
Equating this to Eq.\comcstwo, we find that
\eqn\liusidentity{
\tF_{ij}(k) =
\int {d^{p+1}x\over (2\pi)^{p+1}}
 ~L_*\Big[{\sqrt{\hbox{det}(1-\theta \hat F)}}\,
\big(\hF(1-\theta\hF)^{-1}\big)_{ij}\, W(x, C)\Big] *
e^{ik.x} }

This relates the commutative field strength $F$ to the non-commutative
field strength $\hF$, therefore it amounts to a closed-form expression
for the Seiberg-Witten map. This identity was previously conjectured
starting from a Poisson approximation by Liu\refs\liu, who checked that to
order $(\hA)^3$, the RHS agrees with the $\cO(\hA)^3$ result of
Ref.\refs\mehenwise. Here we see that it follows from the equivalence
of commutative and noncommutative Chern-Simons couplings of a
$p$-brane to a Ramond-Ramond $(p-1)$-form.

\subsec{Coupling to the RR Form $C^{p-3}$ and Lower Forms}

The coupling of a noncommutative D$p$-brane  to the RR form
$C^{(p-3)}$ was written down in Ref.\refs\smnvs\ for the case
of constant RR field:
\eqn\csconstthree{
\half\,\mu_p\,\epsilon^{i_1...i_{p+1}}\,
{\tilde C}^{(p-3)}_{i_1...i_{p-3}}(0)\int
{d^{p+1}x\over (2\pi)^{p+1}} \,
\sqrt{\hbox{det}(1-\theta \hat F)}
\,(Q^{-1})_{i_{p-2}i_{p-1}}
(Q^{-1})_{i_pi_{p+1}} }
where the 2-form $Q^{-1}$ is given in Eq.\qinv. For
spatially varying  $C^{(p-3)}$ we can therefore write the coupling  as:
\eqn\csthree{
\eqalign{
&\half\,\mu_p\,\epsilon^{i_1...i_{p+1}}\,
{\tilde C}^{(p-3)}_{i_1...i_{p-3}}(-k)~\times\cr
&\int
{d^{p+1}x\over (2\pi)^{p+1}} \,
L_*\Big[{\sqrt{\hbox{det}(1-\theta \hat F)}}
\,(Q^{-1})_{i_{p-2}i_{p-1}}
(Q^{-1})_{i_pi_{p+1}} W(x, C)\Big] * e^{ik.x} \cr} }

This is to be compared with the commutative coupling. We do not expect
this to give us new information about the relation between commutative
and noncommutative gauge fields, since the Seiberg-Witten map has
already been obtained in the previous subsection by comparing the
couplings of $C^{(p-1)}$. Therefore, comparing the couplings of
$C^{(p-3)}$ can at best provide a consistency check of what we have
already deduced, unless we have further information about derivative
corrections.

In the DBI approximation of slowly varying fields, the commutative
coupling is:
\eqn\cscommthree{
\half\,\mu_p\,\epsilon^{i_1...i_{p+1}}\,
{\tilde C}^{(p-3)}_{i_1...i_{p-3}}(-k)\int
{d^{p+1}x\over (2\pi)^{p+1}}
\,(B+F)_{i_{p-2}i_{p-1}}
(B+F)_{i_pi_{p+1}} e^{ik.x} }
Inserting Eq.\qinv\ for $Q^{-1}$ in  Eq.\csthree, and comparing with
Eq.\cscommthree, we find that in the DBI approximation we must have:
\eqn\compindbi{
\eqalign{
&\int d^{p+1}k'\, \tF_{ij}(k')\, F_{kl}(k-k') \cr
&=\int
{d^{p+1}x\over (2\pi)^{p+1}} \,
L_*\Big[{\sqrt{\hbox{det}(1-\theta \hat F)}}
\,\Big(\hF(1-\theta\hF)^{-1}\Big)_{ij}
\Big(\hF(1-\theta\hF)^{-1}\Big)_{kl} 
W(x, C)\Big] * e^{ik.x} \cr} }
To arrive at this expression we have made  use of  the identities
Eqs.\idone\ and \liusidentity.

First of all, for strictly constant $F$, the two sides match since in
this case we have
$$
F = \hF(1-\theta\hF)^{-1}
$$
and $\tF(k) \sim \delta(k)$. Then we can pull all the $F$ and $\hF$
out of the integrals, leaving $\delta(k)$ on both sides.

For slowly-varying $F$, we can use a procedure described in
Ref.\refs\smnvs\ and used in Ref.\refs\liu\ where it leads to Eq.(5.8)
of that paper. This consists of the replacement
\eqn\ourproc{
\int {d^{p+1}x \over (2\pi)^{p+1}}
\sqrt{\det(1-\theta\hF)} \rightarrow \int {d^{p+1}X \over (2\pi)^{p+1}} }
which, in the present case, results in the equation:
\eqn\slowvar{
F_{ij}(X(x)) F_{kl}(X(x)) = \Big(\hF(1-\theta\hF)^{-1}\Big)_{ij}
\Big(\hF(1-\theta\hF)^{-1}\Big)_{kl} }
This is just the square of equation (5.8) in Ref.\refs\liu.

This can be extended in a similar way to couplings involving the lower
RR forms. In all these cases, it is interesting to examine how the
derivative corrections match up, a point which we intend to address in
a subsequent work.

\newsec{Inclusion of Transverse Scalars}

In this section we include RR couplings to the scalars
$\hPhi^a$, $a=p+1,\ldots,9$ that represent the transverse degrees of
freedom of the noncommutative brane. One important effect of these
scalars is to modify the Wilson lines by a term depending on an
arbitrary momentum $q_a$. Another source of coupling between these
scalars and the RR field comes about through the noncommutative
analogue of Myers terms\refs\smnvs. 

\subsec{Modification of the Wilson line} 

We start by considering the scalar-dependence of the open Wilson
lines. We will extract some $q$-dependent couplings to transverse
scalars arising from this dependence. One interesting consequence will
be a derivation of the Seiberg-Witten map for transverse scalars.

The open Wilson line including transverse scalars is given by:
\eqn\tsone{
W'(x,C) = P_{*}~\hbox{exp}\left[i\int_0^1 d\tau 
\Big({\partial\xi^i(\tau)\over\partial\tau}\hA_i(x+\xi(\tau)) + 
q_a{\hat \Phi}^a(x+\xi(\tau))\Big)\right] }
Inserting this definition in place of $W(x,C)$ in Eq.\liutwo, and
denoting the LHS by $\cQ'(k)$, one finds that the couplings to a
general spatially varying supergravity mode are:
\eqn\tstwo{
\cQ'(k) = \sum_{m=0}^{\infty}~\int{{d^{p+1}x}\over{(2\pi)^{p+1}}} 
~ \cQ'_m(x) ~ e^{ik.x}}
where the $\cQ'_m$ are given by
\eqn\tsqs{
\eqalign{
\cQ'_m(x) &= {1\over m!}\sum_{k=0}^m {m\choose k}(\theta\del)^{i_1}
\ldots (\theta\del)^{i_k} (iq)_{a_{k+1}}\ldots (iq)_{a_m}~\times\cr
&\langle \cO_1(x),\ldots, \cO_n(x),
\hA_{i_1}(x),\ldots,\hA_{i_k}(x),
\hPhi^{a_{k+1}}(x),\ldots,\hPhi^{a_m}(x)\rangle_{*_{n+m}}\cr} }

For the coupling to the RR top form $C^{p+1}$, the operator
$\sqrt{\det(1-\theta\hF)}$ that must be smeared over the Wilson line
has to be generalised by the addition of terms coming from the
pullback of the RR field components transverse to the brane. We will
return to this in the following subsection. For now we will ignore
such terms and just focus on the $q$-dependence of the coupling. 

Let us plug in this operator into the expansion of the Wilson line of
Eq.\tsone. To the order we are working, we need not consider the terms
beyond $\cQ'_3(x)$. The terms which do not contain any power of $\hat
\Phi$ are exactly the same as the ones considered in the earlier
section. Now let us collect the terms which contain one power of
$\hPhi$ from $\cQ'_1, \cQ'_2$ and $\cQ'_3$.
\eqn\tsonephi{
\eqalign{
\cQ'_1(x) :~& iq_{a_1}\big \{ {\hat\Phi}^{a_1} -
{1\over2}\theta^{kl}\langle \hat F_{lk}, 
{\hat\Phi}^{a_1}\rangle_{*_2} - {1\over4}\theta^{ij}\theta^{kl}\langle 
\hat F_{jk}, \hat F_{li}, \hat \Phi^{a_1}\rangle_{*_3} \cr
&~~~~~~+
{1\over8}\theta^{ij}\theta^{kl}\langle\hat F_{ji}, \hat F_{lk}, \hat
\Phi^{a_1}\rangle_{*_3}\big \} \cr
\cQ'_2(x) :~& iq_{a_1}\big \{\theta^{ij}\partial_j \langle \hat A_i,\hat
\Phi^{a_1}  \rangle_{*_2}- {1\over2}\theta^{ij}\theta^{kl}\partial_j
\langle \hat F_{lk}, \hat A_i, \hat \Phi^{a_1}\rangle_{*_3}\big \} \cr
\cQ'_3(x) :~& iq_{a_1} {1\over
{2!}}\theta^{ij}\theta^{kl}\partial_j\partial_l\langle\hat A_i, \hat
A_k, \hat \Phi^{a_1} \rangle_{*_3}}}
The contribution of these terms can be easily shown to be:
\eqn\tstwophi{
\eqalign{
iq_{a_1}\big \{ \hat \Phi^{a_1} &+ {i\over2}\theta^{kl}\langle\hat
A_l, [\hat A_k, \hat \Phi^{a_1}]\rangle_{*_2} +
{1\over2}\theta^{ij}\theta^{kl} \langle \hat A_i, \partial_l\hat A_j,
\partial_k\hat \Phi^{a_1}\rangle_{*_3} + \theta^{ij}\langle \hat A_i, 
\partial_j\hat\Phi^{a_1}\rangle_{*_2} \cr &-
\theta^{ij} \theta^{kl}\langle \hat A_i, \partial_j\hat A_l,
\partial_k \hat\Phi^{a_1}\rangle_{*_3} + {1\over2}\theta^{ij}\theta^{kl}
\langle\hat A_j, \hat A_l,
\partial_i\partial_k \hat\Phi^{a_1}\rangle_{*_3}\big\}}}

After carrying out the Seiberg-Witten map, we expect that the coupling
written in terms of noncommutative variables should give rise to the
coupling in commutative variables in the DBI approximation, which
is given by:
%
%\mu_p ~ e^{iq_a\Phi^a}\,\epsilon^{i_1...i_{p+1}} ~ {\tilde
%C}^{(p+1)}_{i_1...i_{p+1}}(-k, q) ~ \delta^{(p+1)}(k)}
\eqn\tscomone{
\mu_p\,\epsilon^{i_1\dots i_{p+1}}\Big\{{\tilde C}^{p+1}_{i_1\dots
i_{p+1}}(-k, q)\delta^{(p+1)}(k) + iq_a{\tilde \Phi}^a(k)\,
C^{p+1}_{i_1\dots i_{p+1}}(-k, q) + \dots \Big\} }
This is just the Taylor series expansion of the RR field considered as
a functional of the transverse scalars. Here $\Phi$ is the SW
transform of the noncommutative $\hat\Phi$. Hence the linear coupling
in $\hat\Phi$, after the SW map, should become
$iq_{a_1}\Phi^{a_1}$. 

It follows that the SW map of a transverse scalar, to this order, is
given by the quantity in braces in Eq.\tstwophi\ above.  One can read
off the SW map for $\hat\Phi$ from that of $\hat A_b$
(Ref.\refs\mehenwise) by dimensional reduction\refs\garousitach. It is
easy to show that both of them match exactly to this order as
expected.

\subsec{Noncommutative Myers Terms}

In this section we study the couplings of noncommutative $p$-branes to
RR fields of rank greater than $p+1$. Our strategy will be as
follows. For a collection of $N$ D-instantons, we know the coupling
\refs{\myers,\ramswati\watirev}\ of the non-Abelian scalars $\phi^i$ 
to all the RR $p$-form fields. In this coupling we may substitute
$\phi^i=X^i\equiv x^i+\theta^{ij}\hA_j(x)$, $i=0,\ldots,p$,
representing the classical solution for a noncommutative $p$-brane
along with fluctuations. In this way we find the corresponding
couplings of the noncommutative brane, valid for constant RR
fields. As in the preceding sections of this paper, we then smear
these operators over an open Wilson line to derive the couplings to
spatially varying RR fields. Since we know that a single commutative
$p$-brane has no couplings to RR forms of rank greater than $p+1$, we
will find another interesting identity similar to that in Eq.\idone,
but this time involving the transverse scalars.

For simplicity let us start with the case of a Euclidean D1-brane
coupling to the RR 4-form in type IIB. The coupling of $N$
D-instantons to the RR 4-form is
\eqn\instcoup{
\tr\Big( {1\over 2! 2^2}\left(-i[\phi^{i_1},\phi^{i_2}]\right)
\left(-i[\phi^{i_3},\phi^{i_4}]\right) C^{(4)}_{i_1 i_2 i_3 i_4}
\Big)}
Here $\phi^i$ represent all 10 transverse scalars on a
D-instanton. Now insert $\phi^1=X^1,~\phi^2=X^2$. The remaining
$\phi^i$ are renamed $\hPhi^a$, they represent the scalars
transverse to the noncommutative D1-brane. Thus we find the coupling:
\eqn\finalcoup{
{1\over 2! 2}\,\epsilon_{ij}\,\tr\Big(
\big(-i[X^i,X^j]\big) \big(-i[\hPhi^a,\hPhi^b]\big)
- \big(-i[X^i,\hPhi^a]\big) \big(-i[X^j,\hPhi^b]\big)\Big)
C^{(4)}_{12ab} }
Making the replacements
\eqn\repl{
\eqalign{
-i[X^1,X^2] &= Q^{12} = \theta^{12}(1+\theta^{12}\hF_{12})\cr 
[X^i,\hPhi^a] &= i\theta^{ij} D_j\hPhi^a\cr}}
the operator turns into:
\eqn\finalcouptwo{
\theta^{12}\Big( (1+\theta^{12}\hF_{12})\big(-i[\hPhi^a,\hPhi^b]\big)
+ \theta^{ij}D_j\hPhi^a D_i\hPhi^b \Big) }
Now this can be smeared over the Wilson line as in Eq.\nonccoup\ to
find the operator coupling to $C_{12ab}(-k)$.  The
result should be compared with the corresponding coupling on a
commutative brane. However, it is well-known that a single commutative
$p$-brane does not couple to forms of rank greater than $p+1$, so the
expression that we obtain must be equal to zero. As a result we find
that: 
\eqn\wtfinalcoup{
0= \int {d^{p+1}x\over (2\pi)^{p+1}} L_*\Big[
\Big((1+\theta^{12}\hF_{12})\big(-i[\hPhi^a,\hPhi^b]\big)
+ \theta^{ij}D_j\hPhi^a D_i\hPhi^b\Big)W'(x,C)\Big] * e^{ik.x} }
This is a new identity. An explicit proof of this, for the case
where $q_a=0$ in Eq.\tsone\ (and hence $W'(x,C)=W(x,C)$) is given 
in Appendix B.

A more general version of the above identity can be obtained by
inserting into the coupling Eq.\instcoup\ the classical solution and
fluctuations for a system of $n$ D$p$-branes constructed out of
infinitely many D-instantons. In this case, the commutative branes are
non-Abelian and therefore they do couple to the higher RR forms. These
known couplings have to be equal to the noncommutative couplings
obtained by following through the above procedure. Our identity
stating that Eq.\wtfinalcoup\ vanishes will then arise as the special
case for $n=1$.

\newsec{Non-BPS branes and RR couplings}

It is well-known that non-BPS branes in superstring theory also couple
to RR forms. In commutative variables, these couplings for a single
non-BPS brane are given by:
\eqn\nbcomcs{
\hat S_{CS} = {{\mu_{p-1}}\over {2T_0}}\int~dT\wedge \sum_n
C^{(n)}\wedge ~ e^{B+F}}
where $T$ is the tachyon field and $T_0$ is its value at the minimum
of the tachyon potential. (More general tachyon couplings have been
found in Refs.\refs{\kenwilk,\kmm,\kral,\ttu}, but here
we will only deal here with the term linear in $T$). In this section,
we would like to express the couplings of RR fields to a non-BPS brane
in a constant B-field, in terms of noncommutative variables in the
background independent $\Phi = -B$ description.

In a previous paper\refs\smnvs, we found these couplings for the case
of constant RR fields. Here we generalize them to non-constant RR
fields. As we have seen above for BPS D-branes, RR forms couple to
gauge invariant operators in the noncommutative world-volume gauge
theory of the D-brane. These operators are obtained by smearing the
operators which couple to constant RR fields, over a straight Wilson
line. Here we follow the same prescription for non-BPS branes. It is
important to note that there is no direct matrix-theory derivation of
this prescription in this case. We will be able to show that it
nevertheless gives rise to the correct commutative couplings, which is
strong a posteriori justification for it.

Consider the coupling of a Euclidean non-BPS D$p$-brane
with an even number of world-volume directions, to the RR form
$C^{(p)}$, in the commutative description:
\eqn\nbcomcsone{
\eqalign{
{{\mu_{p-1}}\over {2T_0}}\int dT\wedge C^{(p)} &= {{\mu_{p-1}}\over
{2T_0}} \int {d^{p+1}x}
\,\epsilon^{i_1 i_2\ldots i_{p+1}}\,\partial_{i_1}T(x)~
C^{(p)}_{i_2 \ldots i_{p+1}}(x) \cr
&={{\mu_{p-1}}\over {2T_0}}\int {d^{p+1}k}~
\epsilon^{i_1\ldots i_{p+1}}(-ik_{i_1})\,{\tilde T}(k)\,\tilde
C^{(p)}_{i_2\ldots i_{p+1}}(-k)}}

In Ref.\refs\smnvs\ it was argued that the noncommutative
generalisation of this coupling, for constant RR fields, is
\eqn\nbnccszero{
{{\mu_{p-1}}\over {2T_0}}\,
\epsilon^{i_1i_2\ldots i_{p+1}}\,
{\tilde C}^{(p)}_{i_2\ldots i_{p+1}}(0)\int 
{d^{p+1}x\over (2\pi)^{p+1}}
\,\sqrt{\det(1-\theta \hat F)}
\,{\cal D}_{i_1}\hT(x)}
where
\eqn\dit{
{\cal D}_i\hT(x) = -i~Q^{-1}_{ij}[X^j, \hT(x)] }
Then, the same RR form $C^{(p)}$ couples to a noncommutative non-BPS
D$p$-brane through the following coupling for each momentum mode:
\eqn\nbnccsone{
{{\mu_{p-1}}\over {2T_0}}\,\epsilon^{i_1 i_2\ldots i_{p+1}}\,
{\tilde C}^{(p)}_{i_2 \ldots i_{p+1}}(-k)\int 
{d^{p+1}x\over (2\pi)^{p+1}}
\,L_{*}\Big[{\sqrt{\hbox{det}(1-\theta \hat F)}}
\,{\cal D}_{i_1}\hT(x)\, W(x, C)\Big] * e^{ik.x}}

Now we will show that this coupling is identical to Eq.\nbcomcsone\ on
carrying out the SW map, to terms containing at most three open-string
fields. For this, we need the SW map of the quantity $\partial_a T$
(which has previously been examined in Ref.\garousitach, to second
order in open-string fields).  This can be read off
straightforwardly from the SW map of $F$, and is given by:
\eqn\swoft{
\eqalign{
\partial_a T =&~ [C_a, \hT] + \theta^{ij}\langle \hat A_i, [C_a,
\hT]\rangle_{*_2} + {1\over 2}\theta^{ij}\langle \hat F_{ij}, [C_a,
\hT] \rangle_{*_2} + \theta^{ij}\langle \hat F_{ai}, [C_j,
\hT]\rangle_{*_2} \cr 
&~+ {1\over2}\theta^{ij}\theta^{kl}\partial_i
\partial_k \langle [C_a, \hT], \hat A_l, \hat A_j \rangle_{*_3} -
{1\over2} \theta^{ij}\theta^{kl} \partial_k \langle \hat F_{ij}, [C_a,
\hT], \hat A_l \rangle_{*_3} \cr
&~ - \theta^{ij}\theta^{kl}\partial_k
\langle \hat F_{ai}, [C_j, \hT], \hat A_l \rangle_{*_3} 
+ {1\over 2}
\theta^{ij}\theta^{kl} \langle \hat F_{ai}, [C_j, \hT], \hat F_{kl}
\rangle_{*_3} \cr
&~ + {1\over8}\theta^{ij}\theta^{kl} \langle[C_a, \hT],
\hat F_{ij}, \hat F_{kl} \rangle_{*_3} + {1\over4} \theta^{ij}
\theta^{kl} \langle [C_a, \hT], \hat F_{jk}, \hat F_{il}
\rangle_{*_3}  \cr
&~ + \theta^{ij}\theta^{kl} \langle \hat F_{ik}, \hat
F_{al}, [C_j, \hT]\rangle_{*_3}}}
where $C_a = -i \theta^{-1}_{ab} X^b$.

Now let us expand the coupling on the noncommutative side. For this
we expand the operator that is smeared over the Wilson line in 
Eq.\nbnccsone\ to terms with one $\hT(x)$ and at most two $\hat F$'s: 
\eqn\opone{
\eqalign{
{\cal O}_i(x) \equiv &~ {\sqrt{\det(1-\theta \hat F)}}
\,(-i)\, Q^{-1}_{ij}[X^j, \hT(x)] \cr
=&~ {\sqrt{\det(1-\theta \hat F)}}\left\{ [C_i, \hT] + \left({{\hat F}
\over {1-\theta\hat F}}\right)_{ik}\theta^{kl}\,[C_l, \hT] \right\} \cr
=&~ \left\{ 1 - {1\over2}\tr (\theta\hat F)
-{1\over4}\tr (\theta\hat F\theta\hat F) +
{1\over8}\big(\tr (\theta\hat F)\big)^2\right\}[C_i, \hT] \cr
&~ -{1\over2}\tr (\theta\hat F)\,\hat F_{ik}\theta^{kl}\,[C_l, \hT] 
+ \hat F_{ik}\theta^{kl}\,[C_l, \hT] + \hat F_{ij}\theta^{jk}\hat
F_{kl}\theta^{lm}\, [C_m, \hT]}}
With this expansion at hand, it is straightforward to show that the
noncommutative coupling is equivalent to the commutative one using the 
SW map in Eq.\swoft.

Clearly one can extend this logic to obtain the coupling of a
noncommutative non-BPS $p$-brane to lower and higher RR forms, though
we will not work this out here. However, it is remarkable that the
prescription formulated for BPS branes works for non-BPS branes in the
case we have investigated. This fact might provide a clue to the open
problem of constructing unstable non-BPS D-branes from matrix theory.
Various results on the construction of brane-antibrane pairs from
matrix theory can be found in Refs.\refs{\krs,\miao,\mandalwadia}.

\newsec{Conclusions}

We have seen that the computation of RR couplings for noncommutative
branes, initiated in Ref.\smnvs, can be elegantly extended to
spatially varying RR fields using the ideas in
Refs.\refs{\liu,\dastrivedi}. Comparison of the noncommutative
couplings to commutative ones gives rise to a number of interesting
identities including the Seiberg-Witten map.

These ideas were also extended to incorporate transverse scalars, with
analogous results. The generalisation to RR couplings of unstable,
non-BPS branes also gives sensible results despite the fact that in
this case the construction of  the branes starting from matrix theory
is on less solid ground.

Some interesting problems that we have not addressed include the
question of whether one can gain some insight into the nonabelian SW
map by these methods. One should also ask what interesting physical
effects follow from noncommutative RR couplings, analogous for example
to the Myers effect for nonabelian branes.
\bigskip\medskip

\noindent{\bf Acknowledgements}

We are grateful to Sumit Das for initial collaboration, and are happy
to acknowledge helpful discussions with Sumit Das, Shesansu Pal and
Sandip Trivedi. One of us (S.M.) would like to acknowledge the ITP,
Santa Barbara, for hospitality during the completion of this
work. This research was supported in part by the National Science
Foundation of the U.S. under Grant No. PHY99-07949.

\appendix{A}{Proof of Eq.\idone\ for $\theta$ of Rank Two}

In this appendix we present a proof of the identity \idone\ for the
case where $\theta$ is of rank two, namely a Euclidean D1-brane.  For
this, we write down all terms of ${\cal O}(\hat A^n)$ and show that
for any $n>0$, the sum of the contributions vanishes identically.

In the rank two case, the operator ${\sqrt{\hbox{det}(1-\theta \hat
F)}}$ just becomes $1 + \theta^{12}\hat F_{12}$. Recall that the
quantities $\cQ_n$ defined in Eq.\liumore\ contain contributions of
order $\hA^n$ from the expansion of the Wilson line. In addition we
can select terms  of order $0,1,2$ in $\hA$ from  the operator
$(1+\theta^{12}\hF_{12})$. Thus, terms of  order $\hA^n$ can arise from
$\cQ_n(x)$, $\cQ_{n-1}(x)$ and $\cQ_{n-2}(x)$. These terms are:
\eqn\qns{
\eqalign{
{\rm From}~\cQ_n(x):~&
{1\over{n!}}\theta^{i_1j_1}...\theta^{i_nj_n}
\partial_{j_1}...\partial_{j_n}\langle\hat A_{i_1},..., \hat
A_{i_n}\rangle_{*_n} \cr
{\rm From}~\cQ_{n-1}(x):~&
{1\over {(n-1)!}}\theta^{i_1j_1}...\theta^{i_{n-1}j_{n-1}}
\partial_{j_1}...\partial_{j_{n-1}}\langle\theta^{12}(\partial_1\hat
A_2 - \partial_2\hat A_1), \hat A_{i_1},..., \hat
A_{i_{n-1}}\rangle_{*_n} \cr
{\rm From}~\cQ_{n-2}(x):~&
{1\over {(n-2)!}}\theta^{i_1j_1}...\theta^{i_{n-2}j_{n-2}} 
\partial_{j_1}...\partial_{j_{n-2}}\langle -i\theta^{12}[\hat A_1,
\hat A_2], \hat A_{i_1},..., \hat A_{i_{n-2}}\rangle_{*_{n-1}}}}
The summed indices in the above expression can only take the values 1
and 2. Writing them explicitly, we find that the above contributions 
are:
\eqn\qnsagain{
\eqalign{
{\rm From}~\cQ_n(x):~&
{(\theta^{12})}^n\sum_{r=0}^n {{(-1)^r}\over{r!(n-r)!}}
\partial_1^r\partial^{n-r}_2 \langle \hat A_1^{n-r}, 
\hat A_2^{r}\rangle_{*_n} \cr
{\rm From}~\cQ_{n-1}(x):~&
{(\theta^{12})}^n\big\{\sum_{q=0}^{n-1}{{(-1)^q}\over{(q+1)!(n-q-1)!}}
\partial_1^q \partial_2^{n-q-1} \langle \hat A_1^{n-q-1},
\partial_1 (\hat A_2^{q+1}) \rangle_{*_n} \cr
&~~~~~~ - \sum_{t=0}^{n-1}{{(-1)^t}\over {t!(n-t)!}}\partial_1^t
\partial_2^{n-t-1} \langle \partial_2(\hat A_1^{n-t}), \hat A_2^t 
\rangle_{*_n}\big\} \cr
{\rm From}~\cQ_{n-2}(x):~&
{(\theta^{12})}^n\big\{\sum_{p=0}^{n-2}{{(-1)^p}\over{(p+1)!(n-p-1)!}} 
\partial_1^p\partial_2^{n-p-2} \langle \partial_1(\hat A_1^{n-p-1}),
\partial_2 (\hat A_2^{p+1}) \rangle_{*_n} \cr
&~~~~~~ - \sum_{s=0}^{n-2}{{(-1)^s}\over {(s+1)!(n-s-1)!}}\partial_1^s
\partial_2^{n-s-1} \langle \partial_2(\hat A_1^{n-s-1}),
\partial_1(\hat A_2^{s+1}) \rangle_{*_n} \big\}}}
Here we have also used the recursive relation 
\eqn\recurs{
\theta^{ij}\partial_j\langle f_1, \dots, f_n,
\partial_ig\rangle_{\star_n} = i\sum_{j=1}^{n-1}\langle f_1, \dots, 
[f_j, g], \dots, f_{n-1}\rangle_{\star_{n-1}}}
to convert the $*_{n-1}$ products in $\cQ_{n-2}(x)$ to $*_n$ products. 

Now it is easy to check that the three contributions above 
add up to zero. Therefore we have proved the proposed identity 
Eq.\idone\ in the case where  $\theta^{ij}$ is of rank two.

\appendix{B}{Proof of Eq.\wtfinalcoup\ for $\theta$ of Rank Two}

This proof is similar in spirit to the one in the preceding Appendix.
The main difference is that the operator to be smeared over the open Wilson
line is the one in Eq.\finalcouptwo. To simplify the proof we take
$q_a=0$ in Eq.\tsone, so that the $\cQ'_m(x)$ in Eq.\tsqs\ reduce to
the $\cQ_m(x)$ in Eq.\liumore.

As before, we collect all the terms containing $n$ powers of $\hA$,
and they can arise from $\cQ_n,\cQ_{n-1}$ and $\cQ_{n-2}$. Thus we
have the following three terms:
\eqn\qsone{
\eqalign{
{\rm From}~\cQ_n:~&{1\over{n!}}\theta^{i_1j_1}\dots\theta^{i_nj_n}
\partial_{j_1}\dots\partial_{j_n}\Big\{ \langle -i[\hat\Phi^a,
\hat\Phi^b], \hat A_{i_1},\dots, \hat
A_{i_n}\rangle_{\star_{n+1}}\cr 
& + \theta^{ij}\langle
\partial_j\hat\Phi^a,\partial_i\hat\Phi^b,\hat A_{i_1},\dots, 
\hat A_{i_n}\rangle_{\star_{n+2}}\Big\} \cr
{\rm From}~\cQ_{n-1}:~&{1\over{(n-1)!}}
\theta^{i_1j_1}\dots\theta^{i_{n-1}j_{n-1}}
\partial_{j_1}\dots\partial_{j_{n-1}}\cr
&\Big\{ \langle -i[\hat\Phi^a,
\hat\Phi^b], \theta^{12}(\partial_1\hA_2 - \partial_2\hA_1),
\hat A_{i_1},\dots, 
\hat A_{i_{n-1}}\rangle_{\star_{n+1}}\cr
&+\theta^{ij}\langle\partial_j\hat\Phi^a, -i[\hA_i, \hat\Phi^b], 
\hat A_{i_1},\dots, \hat A_{i_{n-1}}\rangle_{\star_{n+1}}\cr
&+\theta^{ij}\langle -i[\hA_j, \hat\Phi^a],\partial_i\hat\Phi^b,
\hat A_{i_1},\dots, \hat A_{i_{n-1}}\rangle_{\star_{n+1}}\Big\}\cr
{\rm From}~\cQ_{n-2}:~& {1\over{(n-2)!}}
\theta^{i_1j_1}\dots\theta^{i_{n-2}j_{n-2}}
\partial_{j_1}\dots\partial_{j_{n-2}}\cr
&\Big\{ \langle
-i[\hat\Phi^a, \hat\Phi^b], -i\theta^{12}[\hA_1, \hA_2],
\hat A_{i_1},\dots, \hat A_{i_{n-2}}\rangle_{\star_{n}}\cr
& + \theta^{ij}\langle -i[\hA_j, \hat\Phi^a], -i[\hA_i, \hat\Phi^b],
\hat A_{i_1},\dots, \hat A_{i_{n-2}}\rangle_{\star_{n}}}}
Again, these expressions can be simplified using the fact that the summed
indices take only the values 1 and 2:
\eqn\qstwoone{
\eqalign{
{\rm From}~\cQ_n:~&{(\theta^{12})^n\over{n!}} 
\sum_{r=0}^{n}(-1)^r\, {n\choose r}
\partial_1^r\,\partial_2^{n-r}\Big\{\langle -i[\hat\Phi^a, \hat\Phi^b],
\hA_1^{n-r}, \hA_2^r \rangle_{\star_{n+1}} \cr
&+\theta^{12}\langle\partial_2\hat\Phi^a, \partial_1\hat\Phi^b, 
\hA_1^{n-r}, \hA_2^r \rangle_{\star_{n+2}} 
- \theta^{12}\langle\partial_1\hat\Phi^a, \partial_2\hat\Phi^b, 
\hA_1^{n-r}, \hA_2^r \rangle_{\star_{n+2}} \Big\} \cr}}
\eqn\qstwotwo{
\eqalign{
{\rm From}~\cQ_{n-1}:~&{(\theta^{12})^n\over{(n-1)!}} 
\sum_{q=0}^{n-1}(-1)^q\,
{{n-1}\choose q}\partial_1^q\,\partial_2^{n-q-1}\cr
&\Big\{\langle
-i[\hat\Phi^a, \hat\Phi^b], (\partial_1 \hA_2 - \partial_2\hA_1),
\hA_1^{n-q-1}, \hA_2^q \rangle_{\star_{n+1}}\cr 
&+\langle\partial_2\hat\Phi^a, -i[\hA_1, \hat\Phi^b],
\hA_1^{n-q-1}, \hA_2^q \rangle_{\star_{n+1}} -
\langle\partial_1\hat\Phi^a, -i[\hA_2, \hat\Phi^b],
\hA_1^{n-q-1}, \hA_2^q \rangle_{\star_{n+1}}\cr
&+\langle -i[\hA_2, \hat\Phi^a],
\partial_1\hat\Phi^b, \hA_1^{n-q-1}, \hA_2^q \rangle_{\star_{n+1}} -
\langle -i[\hA_1, \hat\Phi^a],
\partial_2\hat\Phi^b, \hA_1^{n-q-1}, \hA_2^q 
\rangle_{\star_{n+1}}\Big\}\cr}}
\eqn\qstwothree{
\eqalign{
{\rm From}~\cQ_{n-2}:~&{(\theta^{12})^{n-1}\over{(n-2)!}} 
\sum_{p=0}^{n-2}(-1)^p\,
{{n-2}\choose p}\partial_1^p\partial_2^{n-p-2}\cr
&\Big\{\langle -i[\hat\Phi^a, \hat\Phi^b], -i[\hA_1,\hA_2],
\hA_1^{n-p-2}, \hA_2^p \rangle_{\star_{n}}\cr
& + \langle -i[\hA_2, \hat\Phi^a], -i[\hA_1,\hat\Phi^b],
\hA_1^{n-p-2}, \hA_2^p \rangle_{\star_{n}}\cr
& - \langle -i[\hA_1, \hat\Phi^a], -i[\hA_2, \hat\Phi^b],
\hA_1^{n-p-2}, \hA_2^p \rangle_{\star_{n}} \Big\}}}

Finally, we rewrite these expressions using the recursion
relation Eq.\recurs, to get:
\eqn\qsthreeone{
\eqalign{
{\rm From}~\cQ_n:~& -i (\theta^{12})^n
\sum_{r=0}^n{{(-1)^r}\over{r!\,(n-r)!}}
\partial_1^r\partial_2^{n-r}\langle[\hat\Phi^a, \hat\Phi^b], 
\hA_1^{n-r}, \hA_2^r \rangle_{\star_{n+1}}\cr
&+(\theta^{12})^{n+1}\sum_{s=0}^n{{(-1)^s}\over{s!\,(n-s)!}}
\partial_1^s\partial_2^{n-s}\big\{\langle\partial_2\hat\Phi^a, 
\partial_1\hat\Phi^b,  
\hA_1^{n-s}, \hA_2^s \rangle_{\star_{n+2}} \cr
& - \langle\partial_1\hat\Phi^a, \partial_2\hat\Phi^b,  
\hA_1^{n-s}, \hA_2^s \rangle_{\star_{n+2}}\big\} }}
\eqn\qsthreetwo{
\eqalign{
{\rm From}~\cQ_{n-1}:~& -i (\theta^{12})^n
\Big\{\sum_{a=0}^{n-1}{{(-1)^a}\over
{a!\,(n-a-1)!}} \partial_1^a\partial_2^{n-a-1}\Big(\langle
\partial_2\hat\Phi^a, [\hA_1, \hat\Phi^b],
\hA_1^{n-a-1}, \hA_2^a \rangle_{\star_{n+1}} \cr
& - \langle [\hA_1, \hat\Phi^a],
\partial_2\hat\Phi^b, \hA_1^{n-a-1}, \hA_2^a
\rangle_{\star_{n+1}}\Big)\cr
& -(i\theta^{12})\sum_{b=0}^{n-1}{{(-1)^b}\over
{(b+1)!\,(n-b-1)!}} \partial_1^{b+1}\partial_2^{n-b-1}\Big(
\langle\partial_1\hat\Phi^a, \partial_2\hat\Phi^b,  
\hA_1^{n-b-1}, \hA_2^{b+1} \rangle_{\star_{n+2}}\cr
& - \langle\partial_2\hat\Phi^a, \partial_1\hat\Phi^b,  
\hA_1^{n-b-1}, \hA_2^{b+1} \rangle_{\star_{n+2}}\Big)\cr
& + \sum_{c=0}^{n-1}{{(-1)^c}\over
{(c+1)!\,(n-c-1)!}} \partial_1^{c}\partial_2^{n-c-1}\Big(
\langle\partial_1[\hat\Phi^a, \hat\Phi^b], 
\hA_1^{n-c-1}, \hA_2^{c+1} \rangle_{\star_{n+1}}\cr
& + \langle[\hat\Phi^a, \hat\Phi^b], 
\hA_1^{n-c-1}, \partial_1(\hA_2^{c+1}) \rangle_{\star_{n+1}}\Big)\cr
& +\sum_{d=0}^{n-2}{{(-1)^d}\over
{(d+1)!\,(n-d-2)!}} \partial_1^d\partial_2^{n-d-1}\Big(\langle
\partial_1\hat\Phi^a, [\hA_1, \hat\Phi^b],
\hA_1^{n-d-2}, \hA_2^{d+1} \rangle_{\star_{n+1}} \cr
& - \langle [\hA_1, \hat\Phi^a],
\partial_1\hat\Phi^b, \hA_1^{n-d-2}, \hA_2^{d+1}
\rangle_{\star_{n+1}}\Big)\cr
& - \sum_{e=0}^{n-1}{{(-1)^e}\over
{e!\,(n-e)!}} \partial_1^e\partial_2^{n-e-1}\langle
[\hat\Phi^a, \hat\Phi^b], \partial_2(\hA_1^{n-e}), \hA_2^e
\rangle_{\star_{n+1}}\Big\}}}
\eqn\qsthreethree{
\eqalign{
{\rm From}~\cQ_{n-2}:~& i(\theta^{12})^n\Big\{
\sum_{f=0}^{n-2}{{(-1)^f}\over
{(f+1)!\,(n-f-1)!}} \partial_1^{f}\partial_2^{n-f-1}\Big(
\langle\partial_1[\hat\Phi^a, \hat\Phi^b], 
\hA_1^{n-f-1}, \hA_2^{f+1} \rangle_{\star_{n+1}}\cr
& + \langle[\hat\Phi^a, \hat\Phi^b], 
\hA_1^{n-f-1}, \partial_1(\hA_2^{f+1}) \rangle_{\star_{n+1}}\Big)\cr
& - 
\sum_{h=0}^{n-2}{{(-1)^h}\over
{(h+1)!\,(n-h-2)!}} \partial_1^h\partial_2^{n-h-1}\Big(\langle
[\hA_1, \hat\Phi^a], \partial_1\hat\Phi^b, 
\hA_1^{n-h-2}, \hA_2^{h+1} \rangle_{\star_{n+1}} \cr
& - \langle \partial_1\hat\Phi^a, [\hA_1, \hat\Phi^b],
\hA_1^{n-h-2}, \hA_2^{h+1}
\rangle_{\star_{n+1}}\Big)\cr 
& - \sum_{i=0}^{n-2}{{(-1)^i}\over
{(i+1)!\,(n-i-1)!}} \partial_1^{i+1}\partial_2^{n-i-2}
\Big(
\langle\partial_2[\hat\Phi^a, \hat\Phi^b], 
\hA_1^{n-i-1}, \hA_2^{i+1} \rangle_{\star_{n+1}}\cr
& + \langle[\hat\Phi^a, \hat\Phi^b], 
\hA_1^{n-i-1}, \partial_2(\hA_2^{i+1}) \rangle_{\star_{n+1}}\Big)\cr
& +
\sum_{k=0}^{n-2}{{(-1)^k}\over
{(k+1)!\,(n-k-2)!}} \partial_1^{k+1}\partial_2^{n-k-2}\Big(\langle
[\hA_1, \hat\Phi^a], \partial_2\hat\Phi^b, 
\hA_1^{n-k-2}, \hA_2^{k+1} \rangle_{\star_{n+1}} \cr
& - \langle \partial_2\hat\Phi^a, [\hA_1, \hat\Phi^b],
\hA_1^{n-k-2}, \hA_2^{k+1}
\rangle_{\star_{n+1}}\Big) \Big\}
 }}

The reader will readily verify that the three expressions
in Eqs.\qsthreeone, \qsthreetwo\ and \qsthreethree\ above add up to
0. This completes the proof.

\listrefs
\end